\documentstyle[epsf,12pt]{article}

\makeatletter

 \@addtoreset{equation}{section}
\makeatother

\begin{document}
\topmargin 0pt
\oddsidemargin 5mm

\newcommand {\beq}{\begin{eqnarray}}
\newcommand {\eeq}{\end{eqnarray}}
\newcommand {\non}{\nonumber\\}
\newcommand {\eq}[1]{\label {eq.#1}}
\newcommand {\defeq}{\stackrel{\rm def}{=}}
\newcommand {\gto}{\stackrel{g}{\to}}
\newcommand {\hto}{\stackrel{h}{\to}}

\newcommand {\1}[1]{\frac{1}{#1}}
\newcommand {\2}[1]{\frac{i}{#1}}

\newcommand {\th}{\theta}
\newcommand {\thb}{\bar{\theta}}
\newcommand {\ps}{\psi}
\newcommand {\psb}{\bar{\psi}}
\newcommand {\ph}{\varphi}
\newcommand {\phs}[1]{\varphi^{*#1}}
\newcommand {\sig}{\sigma}
\newcommand {\sigb}{\bar{\sigma}}
\newcommand {\Ph}{\Phi}
\newcommand {\Phd}{\Phi^{\dagger}}
\newcommand {\Sig}{\Sigma}
\newcommand {\Phm}{{\mit\Phi}}
\newcommand {\eps}{\varepsilon}
\newcommand {\del}{\partial}
\newcommand {\dagg}{^{\dagger}}
\newcommand {\pri}{^{\prime}}
\newcommand {\prip}{^{\prime\prime}}
\newcommand {\pripp}{^{\prime\prime\prime}}
\newcommand {\prippp}{^{\prime\prime\prime\prime}}
\newcommand {\delb}{\bar{\partial}}
\newcommand {\zb}{\bar{z}}
\newcommand {\mub}{\bar{\mu}}
\newcommand {\nub}{\bar{\nu}}
\newcommand {\lam}{\lambda}
\newcommand {\lamb}{\bar{\lambda}}
\newcommand {\kap}{\kappa}
\newcommand {\kapb}{\bar{\kappa}}
\newcommand {\xib}{\bar{\xi}}
\newcommand {\Ga}{\Gamma}
\newcommand {\rhob}{\bar{\rho}}
\newcommand {\etab}{\bar{\eta}}
\newcommand {\tht}{\tilde{\th}}
\newcommand {\zbasis}[1]{\del/\del z^{#1}}
\newcommand {\zbbasis}[1]{\del/\del \bar{z}^{#1}}

\newcommand {\vecv}{\vec{v}^{\, \prime}}
\newcommand {\vecvd}{\vec{v}^{\, \prime \dagger}}
\newcommand {\vecvs}{\vec{v}^{\, \prime *}}

\newcommand {\alpht}{\tilde{\alpha}}
\newcommand {\xipd}{\xi^{\prime\dagger}}
\newcommand {\pris}{^{\prime *}}
\newcommand {\prid}{^{\prime \dagger}}
\newcommand {\Jto}{\stackrel{J}{\to}}
\newcommand {\vprid}{v^{\prime 2}}
\newcommand {\vpriq}{v^{\prime 4}}
\newcommand {\vt}{\tilde{v}}
\newcommand {\vecvt}{\vec{\tilde{v}}}
\newcommand {\vecpht}{\vec{\tilde{\phi}}}
\newcommand {\pht}{\tilde{\phi}}

\newcommand {\goto}{\stackrel{g_0}{\to}}
\newcommand {\tr}{{\rm tr}\,}

\newcommand {\GC}{G^{\bf C}}
\newcommand {\HC}{H^{\bf C}}

\newcommand{\vs}[1]{\vspace{#1 mm}}
\newcommand{\hs}[1]{\hspace{#1 mm}}

\setcounter{page}{0}

\begin{titlepage}

\begin{flushright}
KEK-TH-615\\
hep-th/9903174\\
March 1999
\end{flushright}
\bigskip

\begin{center}
{\LARGE\bf
K\"{a}hler Potential for Global Symmetry Breaking
in Supersymmetric Theories}
\vs{10}

\bigskip
{\renewcommand{\thefootnote}{\fnsymbol{footnote}}
{\large\bf Muneto Nitta\footnote{
e-mail: muneto.nitta@kek.jp, nitta@het.phys.sci.osaka-u.ac.jp.\\
Supported in part by the JSPS Research Fellowships.}
}}

\setcounter{footnote}{0}
\bigskip

{\small \it
Department of Physics,
Graduate School of Science, Osaka University,\\
Toyonaka, Osaka 560-0043, Japan\\  and\\
Theory Division,
Institute of Particle and Nuclear Studies, KEK,\\
Tsukuba, Ibaraki 305-0801, Japan
}
\end{center}
\bigskip

\begin{abstract}
We have developed $N=1$ supersymmetric nonlinear realization methods, 
which realize global symmetry breaking in $N=1$ supersymmetric theories. 
The target space of nonlinear sigma models with a linear model origin is 
a $\GC$-orbit, which is a non-compact non-homogeneous K\"{a}hler manifold. 
We show that, if and only if the orbit is open, 
it includes a compact homogeneous K\"{a}hler manifold as a submanifold, 
and a class of strictly $G$-invariant K\"{a}hler potentials reduces to 
a K\"{a}hler potential $G$-invariant up to a K\"{a}hler transformation on 
the submanifold. 
Hence, in the case of an open orbit, 
the most general low-energy effective K\"{a}hler potential 
can be written as the sum of those of the compact submanifolds and 
an arbitrary function of strictly $G$-invariants. 
\end{abstract}

\end{titlepage}

\newpage
\section{Introduction}

Supersymmetric nonlinear sigma models~\cite{Zu,WB} 
have been used in many physical applications, 
such as low-energy effective Lagrangians 
of supersymmetric gauge theories~\cite{SGT,BGS} 
and coset unification models~\cite{CUM}.

In gauge theories, 
the target spaces of nonlinear sigma models are 
the classical (or quantum modified) moduli spaces. 
They are obtained by integrating out gauge fields 
which obtain masses by the Higgs mechanism.  
In supersymmetric theories, this procedure can be understood 
by the method of the K\"{a}hler quotient~\cite{HKLR}.
On the other hand, 
in the case of global symmetry breaking, 
low-energy effective Lagrangians can be obtained by 
integrating out massive fields, such as Higgs fields. 
The target manifolds can be constructed by a method involving 
the nonlinear realization of global symmetry~\cite{CCWZ}. 
(These two procedures just correspond to 
solving the D-term or F-term flatness conditions~\cite{D-flat}.)  
Much progress concerning nonlinear realization 
in supersymmetric theories 
has been made by many authors~\cite{Zu,BKMU}--\cite{SWZ}.\footnote{
For a review of the supersymmetric nonlinear realization, 
see Refs.~\cite{El2,BL,BKY,Ku}. 
Especially, we recommend Ref.~\cite{Ku}.}
A coupling to gauge fields has been discussed in Ref.~\cite{gauging}, 
a coupling to matter fields in Ref.~\cite{BKMU,LRM,matter} and 
supersymmetric Wess-Zumino-Witten terms in Ref.~\cite{SWZ}.

In this paper, we consider global symmetry breaking 
without gauge symmetry, 
and develop a supersymmetric nonlinear realization method.

In general, when a global symmetry $G$ breaks down to its subgroup $H$, 
there appear quasi Nambu-Goldstone (QNG) bosons besides 
ordinary Nambu-Goldstone (NG) bosons. 
The low-energy effective Lagrangian is a nonlinear sigma model, 
whose target space is a K\"{a}hler manifold, 
and is parameterized by NG and QNG bosons. 
(Low-energy theorems of these scattering amplitudes are 
discussed in Refs.~\cite{HNOO,HN}.)
The target space can be written as a complex coset space as 
$\GC/\hat H$, where $\GC$ is a complexification of $G$ and 
$\hat H$ is a complex group which includes a complexification of $H$. 
The K\"{a}hler potentials of 
a $G$-invariant effective Lagrangian are constructed by 
Bando, Kuramoto, Maskawa and Uehara (BKMU)~\cite{BKMU}.
There are three types of $G$-invariant K\"{a}hler potentials: 
A-, B- and C-type.
The A- and C-type K\"{a}hler potentials are strictly $G$-invariant; 
on the other hand, the B-type K\"{a}hler potentials are, in general, 
$G$-invariant up to a K\"{a}hler transformation. 

If there is any QNG boson, the target manifold becomes 
a non-compact non-homogeneous manifold; 
on the other hand, if there is no QNG boson, 
the target manifold is a compact homogeneous manifold.
Itoh, Kugo and Kunitomo (IKK) showed that 
a K\"{a}hler potential of a compact homogeneous K\"{a}hler manifold
can be completely written as a sum of B-type K\"{a}hler potentials. 
(These are called pure realizations.)~\cite{IKK}. 
All compact homogeneous K\"{a}hler manifolds have been 
completely classified in Ref.~\cite{BFR}. 
Moreover, coset unification models are based on 
these compact models~\cite{CUM}. 
Hence, many authors have studied 
compact models~\cite{IKK}--\cite{compact}. 
However, unfortunately, these models have no linear-model origin 
in the case of global symmetry breaking 
(without gauge symmetry)~\cite{Le}.\footnote{
In the case where there is a gauge symmetry, 
we can sometimes construct compact models with linear origins.}  
The target space of global symmetry breaking 
must be a non-compact non-homogeneous manifold~\cite{Le,BL,KS}.

Buchm\"{u}ller and Ellwanger have considered 
models where some central charges in $\hat H$ 
are broken by hand from compact homogeneous models~\cite{BE,El1,El2}.
(We call these models as ``broken center models''.)
These models are non-compact non-homogeneous, and 
have strictly $G$-invariant K\"{a}hler potentials 
besides B-type K\"{a}hler potentials. 
However, it is not known whether they have linear-model origins.

We, thus, consider non-compact non-homogeneous K\"{a}hler manifolds 
which have linear-model origins. 
It is known that such models can have strictly $G$-invariant 
K\"{a}hler potentials~\cite{Le,BL,KS,Ni}.
Thus, there remain a question: 
is there any B-type K\"{a}hler potential 
in a model with a linear-model origin?
In this paper, we give an answer to this question. 
The idea is that we can consider submanifolds of the total target manifold 
and can add K\"{a}hler potentials of the submanifolds 
to the total K\"{a}hler potential. 
We find that, in a model with a linear-model origin, 
B-type K\"{a}hler potentials are strictly $G$-invariant 
and are not independent of A- or C-type invariants. 
Nevertheless, we show that if the orbit is open, 
the K\"{a}hler manifold, $\GC/\hat H$, has 
a compact K\"{a}hler submanifold, $\GC/\tilde H$, 
and B-type K\"{a}hler potentials on $\GC/\hat H$ reduce 
to those on $\GC/\tilde H$; 
also, if the orbit is closed, it does not have 
any compact K\"{a}hler submanifold, 
and B-type K\"{a}hler potentials are still strictly $G$-invariant.
Here, $\tilde H$ is a complex subgroup of $\GC$ and 
includes $\hat H$ as a subgroup, $\hat H \subset \tilde H$; 
it is obtained by changing some broken generators 
in ${\cal \GC}- \hat{\cal H}$ to 
unbroken generators by hand.\footnote{
We use Latin characters as Lie algebras of 
the corresponding Lie groups.}  
(Therefore, $\GC/\hat H$ includes $\GC/\tilde H$ as a submanifold.) 
We can thus add B-type K\"{a}hler potential on compact submanifolds 
to the full K\"{a}hler potential in an open orbit.
Moreover, this open-orbit model coincides with a special class of 
the broken-center models considered 
by Buchm\"{u}ller and Ellwanger~\cite{BE,El1,El2}. 
We, thus, also find a linear origin of special cases of these models.

We can conclude that pure realizations~\cite{IKK} 
are not just mathematical models, 
but are embedded in open orbits with linear-model origins.

\bigskip
This paper is organized as follows. 
In Sec.~2, we show that target spaces of sigma models are 
obtained as $\GC$-orbits of the vacuum. 
They can be classified by the value of $\GC$-invariants. 
To be precise, we treat the $O(N)$ model. 
In this model, there are two kinds of $\GC$-orbits: 
one is a closed orbit, and the other is an open orbit.   
We show that, although these orbits have very similar properties, 
an essential point is different on the both orbits: 
the closed orbit does not have a Borel subalgebra in 
the complex isotropy, $\hat {\cal H}$; 
on the other hand, the open orbit has it. 

In Sec.~3, strictly $G$-invariant K\"{a}hler potentials, 
the A- and C-types, 
are constructed by the method of BKMU. 
Although this section does not have any new result, 
we use the result in Sec.~4.

In Sec.~4, we discuss K\"{a}hler potentials $G$-invariant up to 
a K\"{a}hler transformation, the B-type K\"{a}hler potentials. 
We find that, although they are strictly $G$-invariant and 
are not independent of A- or C-type K\"{a}hler potentials, 
they can be reduced to B-type K\"{a}hler potentials on 
a compact submanifold, if and only if the orbit is open. 

Sec.~5 is devoted to conclusions and discussions.
   
In App.~A, we construct the K\"{a}hler potential 
of the compact homogeneous K\"{a}hler manifold 
$O(N)/O(N-2) \times U(1)$ by using the method of IKK~\cite{IKK}.
Since it is embedded to the open orbit, 
this appendix is used in Sec.~4.

\section{Classification of $\GC$-orbits}

$\GC$-orbits can be characterized by 
the values of $\GC$-invariants composed of fundamental fields. 
In general, depending on their value, 
there exist closed orbits and open orbits. 
As an example, let us consider the $O(N)$ model. 
It has both closed and open orbits. 
The closed orbits of this model 
have been discussed in Refs.~\cite{Ni,HN}. 
In this section we investigate both orbits, 
and find that they have essentially different properties.

\subsection{$\GC$-invariants and $\GC$-orbits}
We first prepare fundamental fields, $\vec{\phi} \in V$, 
which belong to a representation ${\bf N}$ of $G=O(N)$. 
Here, $V$ is a representation space, 
and it is complexified by the supersymmetry: $V={\bf C}^N$. 
Since a superpotential includes only chiral superfields, 
its $G$-invariance leads to $\GC$-invariance. 
Assume that there appear an effective superpotential, 
\beq
 W = g \phi_0 (\vec{\phi}\,^2 - f^2)  \;,
\eeq
where $\phi_0$ is a $\GC$-singlet Lagrange multiplier field, 
which has no D-term and $g$ is a constant. 
Here $f^2$ is also a constant and 
can be taken as being real by a field redefinition of $\phi_0$.
By integrating out $\phi_0$ 
(eliminating $\phi_0$ by its equation of motion), 
we obtain a nonlinear field space, $M$ of $\vec{\phi}$ 
which is embedded in $V$ by a constraint, 
\beq
 \vec{\phi}\,^2 - f^2 =0 \;.
\eeq
Since there is only one $\GC$-invariant and it is fixed, 
the field space $M$ is a manifold with a $\GC$-transitive action, 
namely a $\GC$-orbit:
\beq
 M = \{g \cdot \vec{v}|
 \vec{v}\in V,\;\forall g \in G^{\bf C}\} \;,
\eeq
where $\vec{v} \in V$ is an arbitrary vector 
satisfying $\vec{v}^{\,2}=f^2$.
The complex dimension of $M$ is 
$\dim_{\bf C} M = \dim_{\bf C} V-1 = N-1$, 
since one $\GC$-invariant is fixed.

There exist four types of $\GC$-orbits, 
which can be classified by 
the value of the $\GC$-invariant $f^2$ (see Fig.~1): 
\beq
  && \mbox{I) $f^2 > 0$: closed orbits,} \non
  && \mbox{II) $f^2 < 0$: closed orbits,} \non
  && \mbox{III) $f^2 = 0,\;\vec{\phi}\neq \vec{0}$: an open orbit.}\non
  && \mbox{IV) $f^2 = 0,\;\vec{\phi}=\vec{0}$: a closed orbit,}\nonumber
\eeq
Since orbits I and II can be changed to each other 
by inverting the sign of $f^2$, 
it is sufficient to consider orbits of type I.
Moreover, we do not consider the trivial orbit of type IV. 
We, thus, consider the closed orbits I and the open orbit III.
\begin{figure}
 \epsfxsize=8cm
 \centerline{\epsfbox{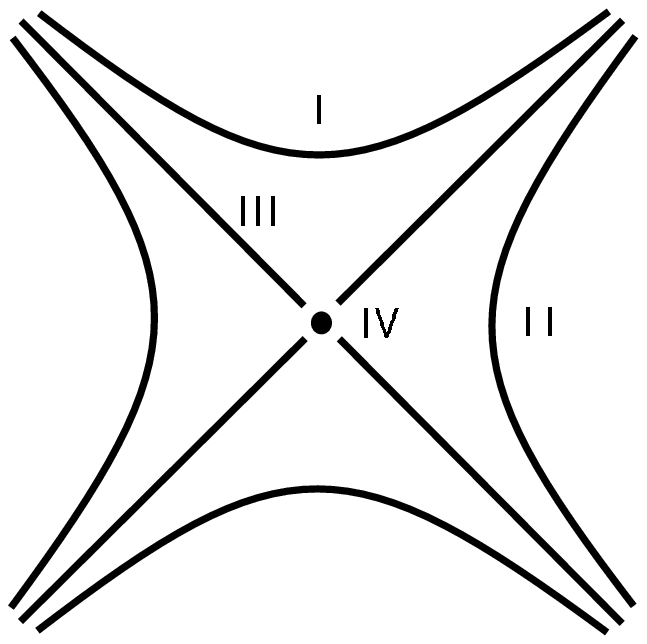}}
 \centerline{\bf Figure\,1}
\begin{footnotesize}
The vertical axis is taken as ${\rm Re}\; V^N$, 
the horizontal axis is taken as ${\rm Im}\; V^{N-1}$ 
and the central point is the origin.
There exist four kinds of orbits.
Orbit I consists of the upper and lower hyperbola 
and orbit II of the left and right hyperbola. 
Although they look separated in this figure, 
they are continuous through other directions of $V$. 
Orbit III looks like a cone, 
and orbit IV is the origin.
\end{footnotesize}
\end{figure}

Since the group action is transitive, 
the orbit can be written as a coset space. 
Namely, if we define the complex isotropy group 
of the vacuum, $\vec{v} = <\vec{\phi}>$, as 
\beq
 \hat H_v = \{g \in G^{\bf C}|g \cdot \vec{v} = \vec{v} \},
\eeq 
the orbit can be written as\footnote{
Here, since the isotropy group at $\vecv = g_0 \vec{v}$ 
can be obtained by the isotropy group at $\vec{v}$ as 
$\hat H_{v\pri} = g_0 \hat H {g_0}^{-1}$, 
we has not written a subscript $v$ below $\hat H$.} 
\beq
 M \simeq \GC/\hat H \;.  \label{coset}
\eeq
The representative of this coset manifold is generated by 
broken generators $Z_i$ in ${\cal \GC} -\hat {\cal H}$ as
\beq
 \xi = \exp (i \Phi \cdot Z) \in \GC/\hat H \;, \label{coset-rep.}
\eeq
where $\Phi^i$ are chiral superfields, 
whose scalar components are the coordinate of $\GC/\hat H$.  

In general, the complex isotropy group is 
larger than a complexification $\HC$ 
of the real isotropy group, 
\beq
 H_v = \{g \in G|g \cdot \vec{v} = \vec{v} \}.
\eeq 
Namely, there is a point such that $\hat H$ can be written as  
\beq
 \hat{\cal H} = \cal \HC \oplus \cal B \;, \label{Hhat-dec.}
\eeq
where $\cal B$ is a nilpotent Lie algebra, 
called a Borel subalgebra in $\cal H$.  
$\cal B$ can be written as 
lower- (or upper-) half triangle matrices in 
a suitable basis~\cite{BKMU}. 
The Borel subalgebra in $\hat{\cal H}$ is an algebra which 
satisfies a commutation relation, 
$[\hat{\cal H}, {\cal B}] \subset {\cal B}$.\footnote{
Therefore $\hat H$ can be written as 
a semidirect product of $\HC$ and $B$: $\hat H= \HC \wedge B$. 
Here the symbol $\wedge$ denotes a semidirect. 
If there are two elements of $\hat H$, 
$hb$ and $h\pri b\pri$, where $h,h\pri \in \HC$ and $b,b\pri \in B$, 
their product is defined as 
$(hb)(h\pri b\pri)= h h\pri (h^{\prime -1} b h\pri) b\pri 
= (h h\pri) (b\prip b)$, 
where $b\prip = h^{\prime -1} b h\pri$~\cite{BKMU,LRM}.}

The complex broken generators $Z_i$ can be classified to two types. 
One is an Hermitian generator, 
and the other is a non-Hermitian generator. 
The latter has a corresponding non-Hermitian unbroken generator, $B_i$, 
in the sense that two Hermitian generators can be composed of 
the linear combinations of $Z_i$ and $B_i$.
(Here $B_i$ need not to be an element of a Borel algebra. 
If $B_i$ is an element of the Borel algebra, 
$Z_i$ and $B_i$ are Hermitian conjugate to each other.) 
The chiral superfields $\Phi^i$ in (\ref{coset-rep.}), 
whose scalar components parameterize $M$, 
are also classified to two types, corresponding to broken generators. 
The chiral superfields, corresponding to Hermitian broken generators, 
are called $mixed$ $types$; on the other hand, 
the chiral superfields, corresponding to non-Hermitian broken generators, 
are called $pure$ $types$. 
These names come from the fact 
that the mixed type includes a QNG boson besides an NG boson 
as a scalar component, 
and the pure type includes two NG bosons without any QNG boson.\footnote{
The fermions in them are sometimes called quasi NG fermions.}

These can be illustrated as follows. 
As stated above, if there are non-Hermitian broken generators, $Z_i$, 
there exist non-Hermitian unbroken generators, $B_i$, 
such that linear combinations of $Z_i$ and $B_i$ are 
Hermitian (broken) generators $X_i$ and ${X_i}\pri$. 
Hence, pure-type generators $Z_i$ in the coset representation 
can be transformed to $a X_i\ + b{X_i}\pri$, 
where $a$ and $b$ are some real constants, 
by a local $\hat H$ transformation from the right of 
the coset representative. 
Since $a X_i+ b{X_i}\pri$ are Hermitian, 
these generate compact directions, corresponding NG bosons. 
Therefore, the scalar components of a pure-type multiplet $\Phi^i$ 
corresponding to $Z_i$ are both NG bosons.

On the other hand, since there is no unbroken partner of 
Hermitian broken generators, 
imaginary parts of scalar components of mixed-type multiplets 
generate non-compact directions, corresponding to QNG bosons.

In the next two subsections, we show how 
these two kinds of broken generators and 
the Borel algebra appear 
in the closed and open orbits, respectively.

\subsection{Closed orbits}
Closed orbits in the $O(N)$ model 
are discussed in Refs.~\cite{Ni,HN}. 
This subsection is devoted to a brief review to them.
The essential feature of closed orbits is 
a supersymmetric vacuum alignment~\cite{KS2,KS,LRM,Ni,HN}. 
(Especially, a geometrical meaning of a vacuum alignment 
has been discussed in our previous paper~\cite{Ni}.) 
By this property, the unbroken real symmetry 
can change from point to point. 

In closed orbits, 
there exists a point $\vec{v}$ such that 
$\hat H$ becomes a reductive group; $\hat H_v = {H_v}^{\bf C}$; 
hence, there is no Borel algebra~\cite{D-flat}.\footnote{
It is a point such that a equation, 
${\del f^2 \over \del {\vec{\phi}}}|_{\vec{\phi} = \vec{v}} 
= C {\del K \over \del {\vec{\phi}}}|_{\vec{\phi}=\vec{v}}$, 
where $C$ is a complex constant and 
$K=\phi\dagg \phi$ is a linear K\"{a}hler potential, 
is satisfied~\cite{D-flat,BGS}. 
An iso-K\"{a}hler surface, 
which looks like a circle in Fig.~1, 
touches the $\GC$-orbit at $\vec{v}$ 
(the nearest point to the origin in Fig.~1).}
Actually, there exist a $\GC$-transformation
such that the vacuum can be transformed to 
\beq
 \vec{v} = \pmatrix{0 \cr \vdots \cr 0 \cr { v}\cr}  \;, 
\eeq
where $v$ is a constant equal to $f$: $v^2 = f^2$. 
We call this point a symmetric point.\footnote{
The points transformed by $G$ from this point 
also satisfy the same property, and are therefore symmetric points. 
It is known that 
the $G$-orbit consisting of symmetric points is unique~\cite{D-flat}. 
Such the $G$-orbit is called D-orbit if we gauge $G$.}
The breaking pattern is $G \to H = O(N-1)$ and 
the broken generators are 
\beq
 X_i 
 = \left(
   \begin{array}{ccc|c}
              &          &          & \vdots  \\
              &\LARGE{0} &          & i         \\ 
              &          &          & \vdots  \\ \hline
      \cdots  &  -i      & \cdots &   
   \end{array}
   \right) \in {\cal G - H} \;\;\;(i = 1,\cdots ,N-1) \;. 
 \label{br.O(N)}
\eeq
A real coset space, $G/H = O(N)/O(N-1)$, is generated by 
these generators. 
Namely, $U \defeq \exp (i \pi \cdot X)$ is a representative of $G/H$, 
where $\pi^i \;(i=1,\cdots,N-1)$ are NG bosons, 
which parameterize $G/H$.
This manifold is a submanifold of the full target manifold $M$, 
and $M$ is just its complexification: 
$M \simeq \GC/\hat H = O(N)^{\bf C}/O(N-1)^{\bf C}$.

The symmetric point is a special point in 
the full target manifold $M$. 
There exist other vacua, transformed by $\GC$ from the symmetric point. 
To consider them, we transform the vacuum by $\GC$ transformation as 
\beq
 { \vec{v}\,\pri  = g_0 \vec{v}} 
 = \pmatrix{0 \cr \vdots \cr 0 \cr -i v \sinh \tht \cr  v \cosh \tht \cr} 
 \defeq \pmatrix{0 \cr \vdots \cr 0 \cr{ \alpha} \cr { \beta}\cr} \;.
\eeq
Here we have put $g_0 \in \GC$ as
\beq
 g_0 
 &=& \exp(i \th X_{N-1}) \non 
 &=& \left(
   \begin{array}{c|cc}               
               \LARGE{1} &\LARGE{0} &          \\ \hline
                         & \cos \th & -\sin\th \\ 
               \LARGE{0} & \sin \th &  \cos\th  
   \end{array}
   \right)
 = \left(
   \begin{array}{c|cc}
               \LARGE{1} & \LARGE{0}   &             \\ \hline
                         & \cosh \tht  & -i\sinh\tht \\ 
               \LARGE{0} & i\sinh \tht &  \cosh\tht  
   \end{array}
   \right) \;,
\eeq
where $\tilde \th = i\th$ is a pure imaginary constant. 
Other $\GC$-transformations are not independent of $g_0$, 
since any vacuum can be reached from the symmetric point 
by using $g_0$ and a $G$-transformation.  
This is because the broken generators at the symmetric point 
belong to a single representation, ${\bf N-1}$, of 
the unbroken symmetry $H$~\cite{Ni}.
We call this vacuum the non-symmetric point. 
The breaking pattern of the global symmetry is $G \to H\pri = O(N-2)$ 
and broken generators are $X_i$ and 
\beq
  {X_i}\pri 
 = \left(
   \begin{array}{ccc|cc}
               &          &      &\vdots&\vdots   \\
               &\LARGE{0} &      & i    & 0 \\ 
               &          &      &\vdots&\vdots   \\ \hline
        \cdots &  -i      &\cdots& 0    & 0 \\ 
        \cdots &   0      &\cdots& 0    & 0  
   \end{array}
   \right)\;(i\pri = 1,\cdots,N-2). \label{Xpri}
\eeq
Although these generators, ${X_i}\pri$, 
were unbroken in the symmetric point, 
they generate newly emerged NG bosons at the non-symmetric point. 
The number of NG bosons is $\dim (G/H\pri) = 2N-3$. 
It has increased more than at the symmetric point. 
Namely, some of QNG bosons at the symmetric point 
have changed to NG bosons, 
and there remains only one QNG boson~\cite{Ni,HN}. 
The unbroken symmetry has been changed there. 
This phenomenon is called ``supersymmetric vacuum 
alignment''~\cite{KS2,KS,LRM,Ni,HN}.

The complex broken generators are transformed as
\beq
 \cases {g_0 X_i g_0^{-1}
             = { {\alpha \over v} {X_i}\pri 
             + {\beta \over v} X_i} \defeq Z_i  \cr
             g_0 X_{N-1} g_0^{-1} = X_{N-1} \defeq Z_{N-1} \;\,
             }  
  \;\;\in {\cal G}^{\bf C}- \hat{\cal H}\pri \;. \label{ge.mixZ}
\eeq
Except for the generator $Z_{N-1}=X_{N-1}$, 
most of broken generators, $Z_i$, become non-Hermitian and, 
thus, are pure-type generators. 
Hence, there are $N_{\rm P}=N-2$ pure-type multiplets and 
the $N_{\rm M}=1$ mixed-type multiplet. 
These are consistent with the numbers of NG and QNG bosons 
counted above. See Table~1. 
(Subscriptions, P and M, 
under $H$-representation of the complex broken generators 
denote pure-types and mixed-types respectively.) 
The complex unbroken generators are also transformed as
\beq
 \cases {g_0 X_i\pri g_0^{-1}
             = { {\beta \over v} {X_i}\pri 
             - {\alpha \over v} X_i} \defeq B_i \cr
             g_0 {H_a}\pri g_0^{-1} = {H_a}\pri \in {\cal H}\pri
             } 
 \;\; \in \hat{\cal H}\pri  \;.\label{ge.mix}
\eeq
There appear non-Hermitian generators $B_i$. 
$B_i$ are partners of $Z_i$ in a sense that, 
from linear combinations of $B_i$ and $Z_i$, 
we can construct Hermitian generators $X_i$ and ${X_i}\pri$. 
$B_i$ does not constitute a Borel subalgebra, 
since $[{\hat{\cal H}}\pri,B_i]\sim B_j$ is not satisfied.
It can be understood from the fact that 
there is no Borel subalgebra at the symmetric point and 
that $\GC$-transformation does not change the commutation relations. 

The target spaces are same 
with being independent of the vacua: $\GC/\hat H \simeq \GC/\hat H\pri$.
\begin{table}
\caption{\bf $O(N)$ with ${\bf N}$, closed orbit}
\begin{center}
\begin{tabular}{|c|c|c|c|c|c|c|}
 \noalign{\hrule height0.8pt}
    & $H$ & $ N_{\rm M}$ & $N_{\rm P}$ & NG & QNG & $H$-sector \\
 \hline
 \noalign{\hrule height0.2pt}
  $v$ & $O(N-1)$ & $N-1$ & $0$   & $N-1$  & $N-1$ &
      $({\bf N-1})_{\rm M}$     \\  
  $\vecv$ & $O(N-2)$ & $1$   & $N-2$ & $2N-3$ & $1$  & 
      $({\bf N-2})_{\rm P} \oplus {\bf 1}_{\rm M}$  \\
 \noalign{\hrule height0.8pt}
 \end{tabular}
 \end{center}
\end{table}

\subsection{Open orbit}
In the last subsection, 
we discussed the closed orbit, characterized by $f^2>0$. 
There was a supersymmetric vacuum alignment, 
and pure-type multiplets appeared at the non-symmetric point. 
In this subsection, we discuss the open orbit characterized by $f^2=0$ 
and show that there is no vacuum alignment. 
Moreover, we find a Borel subalgebra in the complex isotropy, 
differently from the closed orbit.

By a $\GC$-transformation, 
any vacuum on the open orbit can be transformed to 
\beq
 \vec{v} = \pmatrix{0 \cr \vdots \cr 0 \cr iv/\sqrt 2 \cr v/\sqrt 2\cr} \;,  
\eeq
where $v$ can be taken as an arbitrary constant. 
The $G$ transformation can also bring any vacuum to this form, 
but $v$ is not arbitrary. 
In both cases, the breaking patterns are $G \to H= O(N-2)$. 
This coincidence of the breaking patterns means that 
there is no vacuum alignment. 
The number of NG bosons is $\dim(G/H)=2N-3$. 
Besides the Hermitian generators of $O(N-1)$, 
there are additional complex unbroken generators. 
The additional unbroken generators and their broken partners are 
\beq
 B_i
 = \left(
   \begin{array}{ccc|cc}
               &          &      &\vdots&\vdots   \\
               &\LARGE{0} &      & i    & 1 \\ 
               &          &      &\vdots&\vdots   \\ \hline
        \cdots &  -i      &\cdots& 0    & 0 \\ 
        \cdots &  -1      &\cdots& 0    & 0  
   \end{array}
   \right),\;
 Z_i
 = \left(
   \begin{array}{ccc|cc}
               &          &      &\vdots&\vdots   \\
               &\LARGE{0} &      & i    & -1 \\ 
               &          &      &\vdots&\vdots  \\ \hline
        \cdots &  -i      &\cdots& 0    & 0 \\ 
        \cdots &   1      &\cdots& 0    & 0  
   \end{array}
   \right) . \label{non-Her.open}
\eeq
Here, the index $i$ runs over $1$ to $N-2$. 
They can be written as 
$B_i = {X_i}\pri - i X_i$ and $Z_i = {X_i}\pri + i X_i$, 
where $X_i$ and ${X_i}\pri$ 
are given in Eqs.~(\ref{br.O(N)}) and (\ref{Xpri}).  
Hence, by their linear combinations, 
we can obtain Hermitian generators, $X_i$ and ${X_i}\pri$.
The complex broken generators are $Z_i$ and $Z_{N-1}=X_{N-1}$. 
Here, $Z_i$ are pure-type generators 
belonging to a representation ${\bf N-2}$ of $H$ 
and $X_{N-1}$ is a mixed-type generator belonging to a $H$-singlet.
Since all mixed-types are $H$-singlets, 
we can make sure that 
there is no vacuum alignment in the open orbit 
by using the results given in Ref.~\cite{Ni}. 
We summarize these in Table~2.
\begin{table}
\caption{\bf $O(N)$ with ${\bf N}$, open orbit}
\begin{center}
\begin{tabular}{|c|c|c|c|c|c|c|}
 \noalign{\hrule height0.8pt}
    & $H$ & $ N_{\rm M}$ & $N_{\rm P}$ & NG & QNG & $H$-sector \\
 \hline
 \noalign{\hrule height0.2pt}
  $\vec{v}$  & $O(N-2)$ & $1$   & $N-2$ & $2N-3$ & $1$  & 
      $({\bf N-2})_{\rm P} \oplus {\bf 1}_{\rm M}$  \\
 \noalign{\hrule height0.8pt}
 \end{tabular}
 \end{center}
\end{table}

Since $B_i$ satisfy commutation relations, 
$[{\cal H},B_i]\sim B_j$ and $[B_i,B_j]\sim B_k$, 
they are elements of a Borel subalgebra ${\cal B}$ in $\hat{\cal H}$. 
Hence, the target space $M$ can be written as 
$M \simeq \GC / \hat H = O(N)^{\bf C}/O(N-2)^{\bf C} \wedge B$, 
where $B$ denotes a Borel group. 
As stated in Subsec.~2.1, there exist a basis 
such that the Borel subalgebra is represented by 
lower (or upper) half off-diagonal matrices. 
We can thus change the basis to such a basis by  
\beq  
 U = \left(
   \begin{array}{c|cc}               
               \LARGE{1} &\LARGE{0} &          \\ \hline
                         & i/\sqrt 2 & 1/\sqrt 2  \\ 
               \LARGE{0} &-i/\sqrt 2 & 1/\sqrt 2 
   \end{array}
   \right) \;.
\eeq
(In the new basis, $X_{N-1}$ becomes a diagonal matrix.) 
Since $U$ is a unitary matrix, $UU\dagg =U\dagg U=1$, 
the D-term $\vec{\phi}\dagg \vec{\phi}$ does not change.
The vacuum in this basis is
\beq
 U \vec{v} = \pmatrix{0 \cr \vdots \cr 0 \cr v} \;.
\eeq
The complex broken generators, $Z_i$ and $X_{N-1}$, 
and the unbroken generators, $B_i$, 
are represented in the basis as
\beq
&&
U Z_i U\dagg =
 \left(
   \begin{array}{c|ccc|c}   
               &\cdots&   1      &\cdots&        \\ \hline 
               &      &          &      &\vdots  \\ 
               &      &\LARGE{0} &      & -1      \\ 
               &      &          &      &\vdots  \\ \hline
               &      &          &      &        \\ 
   \end{array}
 \right) ,
U X_{N-1} U\dagg =
 \left(
   \begin{array}{c|ccc|c}   
              1&      &          &      &        \\ \hline 
               &      &          &      &        \\ 
               &      &\LARGE{0} &      &        \\ 
               &      &          &      &        \\ \hline
               &      &          &      &-1      \\ 
   \end{array}
 \right) ,\non
&&
U B_i U\dagg =
 \left(
   \begin{array}{c|ccc|c}   
               &      &          &      &        \\ \hline 
         \vdots&      &          &      &        \\ 
             1 &      &\LARGE{0} &      &        \\ 
         \vdots&      &          &      &        \\ \hline
               &\cdots&  -1      &\cdots&
   \end{array}
 \right) \;. \label{tr_basis}
\eeq
Here, we have rearranged an order of the blocks so that 
a $O(N-2)$ part comes to the center.  
They can be summarized as 
\beq
{\cal \GC}-\hat{\cal H}=
 \left(
   \begin{array}{c|c|c}   
              M&\LARGE{P}&         \\ \hline  
               &\LARGE{0}&\LARGE{P}\\ \hline  
               &         &M\\ 
   \end{array}
 \right), 
\hat{\cal H}=
 \left(
   \begin{array}{c|c|c}   
               &                                 & \\ \hline  
 \LARGE{\cal B}&\LARGE{{\cal H}_{\rm SS}^{\bf C}}& \\ \hline  
               &\LARGE{\cal B}                   & \\ 
   \end{array}
 \right) \;,  \label{open-gene.}
\eeq
where P, M, ${\cal B}$ and ${\cal H}_{\rm SS}^{\bf C}$ represent 
the pure-types, the mixed-type, the Borel algebra and 
a semisimple part of $\hat {\cal H}$, namely $O(N-2)$ part, 
respectively.

In the rest of paper, 
we argue the closed orbit in the ordinary basis, 
but the open orbit in the basis changed by the unitary matrix. 
Of course, no result depends on the basis. 
We do not explicitly write $U$ in the open orbit.

\medskip
Here, we give a summary of Sec.~2.
In general, $\GC$-orbits can be characterized 
by the values of $\GC$-invariants. 
There are two types of $\GC$-orbit: 
one is the closed orbit (the orbit I in Fig.~1) and 
the other is the open orbit (orbit III in Fig.~1). 
(Orbit II is a mirror of orbit I and orbit IV is trivial.) 
Both orbits can be written as complex coset spaces as 
$\GC / \hat H = O(N)^{\bf C}/O(N-1)^{\bf C}$ and 
$\GC / \hat H = O(N)^{\bf C}/O(N-2)^{\bf C} \wedge B$, respectively. 
Although these two orbits are topologically 
different near to the origin, 
they become close to each other at infinity, as in Fig.~1. 
Especially, many properties in 
the non-symmetric point on the closed orbit 
and generic points on the open orbit are very similar. 
Namely, the number of pure-type and mixed-type multiplets 
are the same in both cases. 
(See Eqs.~(\ref{ge.mixZ}),(\ref{ge.mix}) and (\ref{non-Her.open}).)
Moreover, their $H$ representations are the same in both cases. 
(Compare the second line of Table~1 and Table~2.) 
However, there is the Borel subalgebra on the open orbit, 
but not on the closed orbit. 
In the next section we show that 
this difference brings essentially distinct results 
concerning to the K\"{a}hler potentials on these orbits.

\section{Strictly $G$-invariant K\"{a}hler potentials}
In this section, 
we construct strictly $G$-invariant K\"{a}hler potentials. 
Although this section does not have any new feature, 
the results are used in the next section. 
Hence, we briefly discuss them.

\subsection{Invariants composed of fundamental fields}
In the $O(N)$-model, there exist one $G$-invariant 
comprising fundamental fields: $\vec{\phi}\dagg \vec{\phi}$. 
This parameterizes a moduli space of global symmetry~\cite{Ni}. 
(We do not discuss this feature.) 
Since this is strictly $G$-invariant, 
a low-energy effective K\"{a}hler potential can be 
written as an arbitrary function of this quantity. 
We showed, in a previous paper~\cite{Ni}, 
that this is just the A-type invariant of BKMU~\cite{Le,BKMU}. 

There is a relation between the fundamental fields 
and the coset representative of $\GC/\hat H$, 
\beq
 \vec{\phi} = \xi \vec{v}|_{\rm F} \;,
\eeq
where F denotes the F-term constraint, 
$\vec{\phi}^2 =f^2$. From this equation, 
the K\"{a}hler potential can be written as
\beq
 K_{\rm A} 
  = f(\vec{\phi}\dagg \vec{\phi})|_{\phi^2 =f^2} 
  = f(\vec{v}\dagg \xi\dagg \xi \vec{v}) \;. \label{A-type}
\eeq
This form does not depend on 
whether the orbit is closed, $f^2>0$, or open, $f^2=0$.

\subsection{Invariants constructed by using projections}
Other strictly $G$-invariants, called as the C-type, 
were found by BKMU~\cite{BKMU}.\footnote{ 
It is not known whether 
these invariants can be written in fundamental fields.} 
We review it here.

Consider projection matrices $\eta_i$, which project $V$ onto 
$\hat H$-invariant subspaces $\eta_i V$. 
These satisfy projection conditions, 
\beq
 \eta\dagg = \eta\,,\; \eta \hat H \eta =\hat H \eta \,,\; \eta^2 = \eta.
  \label{proj}
\eeq
We construct quantities
\beq
 P_i \defeq \xi \eta_i [\xi\dagg \xi]^{-1}_{\eta_i} \eta_i \xi\dagg \;,
\eeq
where $[\cdots]^{-1}_{\eta_i}$ denotes the inverse matrix in 
the $\eta_i$ projected space. 
These transform under $g \in G$ as $P_i \gto g P_i g\dagg$.  
Thus, quantities $\tr (P_i P_j),\;\tr (P_i P_j P_k),\;\cdots$, 
are strictly $G$-invariant.\footnote{
$\tr (P_i) =$ const. is trivial. 
Since ${P_i}^2 = P_i$, we require $i \neq j$ etc.
If two projections satisfy $\eta_i V \subset \eta_j V$ 
then $P_i P_j = P_j P_i = P_i$, 
hence we require $\eta_i V \subset \hspace{-0.3cm} /\;\; \eta_j V$ etc.}  
Hence, a K\"{a}hler potential can be written as 
an arbitrary function of them: 
\beq
 K_{\rm C} = f(\tr (P_i P_j), \cdots) \;\;
 (i \neq j \mbox{ etc.},\; 
  \eta_i V \subset \hspace{-0.4cm} /\;\; \eta_j V \mbox{ etc.}).
\eeq

We next apply this to the $O(N)$ model. 
Projections $\eta$ are different between 
in the closed orbit and in the open orbit. 
We first consider the closed orbit. 
At the symmetric point, 
we can find two projections,\footnote{
At the non-symmetric point, 
as the complex isotropy $\hat H$ transforms to 
$\hat H\pri = g_0 \hat H {g_0}^{-1}$, 
$\eta_i$ to ${\eta_i}\pri = g_0 \eta_i {g_0}^{-1}$. 
Although the first condition in Eq.~(\ref{proj}) is not satisfied, 
it is sufficient to modify $P_i$ as 
${P_i}\pri = \xi\pri {\eta_i}\pri 
[{{\eta_i}\pri}\dagg{\xi\pri}\dagg \xi\pri {\eta_i}\pri]^{-1} 
{{\eta_i}\pri}\dagg {\xi\pri}\dagg$, where $\xi\pri \in \GC/\hat H\pri$. 
Under $g \in G$ transformations, 
${\eta_i}\pri$ also transforms as 
${\eta_i}\pri \to g {\eta_i}\pri g\dagg$.}
\beq
 \eta_1
 = \left(
   \begin{array}{ccc|c}
               &          &      & \\
               &\LARGE{0} &      & \\ 
               &          &      & \\ \hline
               &          &      & 1       
   \end{array}
   \right),\;\; 
 \eta_2
 = \left(
   \begin{array}{ccc|c}
               &          &      & \\
               &\LARGE{1} &      & \\ 
               &          &      & \\ \hline
               &          &      & 0       
   \end{array}
   \right). \label{closed-eta}
\eeq
Hence, there is one C-type invariant on the closed orbit:  
\beq
 K_{\rm C} = f(\tr(P_1 P_2)) \;.
\eeq
Of course, the most general strictly $G$-invariant K\"{a}hler potential 
is an arbitrary function of both the A- and C-type invariants.  

In the open orbit, 
$\hat H$ is the form of 
\beq
 \hat H=
 \left(
   \begin{array}{c|c|c}   
              *&         &     \\ \hline  
              *&\LARGE{*}&     \\ \hline  
              *&\LARGE{*}&*    \\ 
   \end{array}
 \right) \;.
\eeq
Therefore, we can find only one projection, 
\beq
\eta=
 \left(
   \begin{array}{c|c|c}   
              0&         &     \\ \hline  
               &\LARGE{0}&\\ \hline  
               &         &1    \\ 
   \end{array}
 \right)  \;,\label{open-eta}
\eeq
and there is no C-type invariant on the open orbit.\footnote{
$\eta\pri = 
 \left(
   \begin{array}{c|c|c}   
              0&         &     \\ \hline  
               &\LARGE{1}&\\ \hline  
               &         &1    \\ 
   \end{array}
 \right)$ 
also satisfies the projection condition. 
However, from a relation $\eta V \subset \eta\pri V$, 
we can not construct C-type invariants. 
In the case of B-types, 
which is discussed in the next section, 
B-type K\"{a}hler potentials constructed by using $\eta$ and $\eta\pri$ 
are not independent as in the case of pure realizations~\cite{IKK}. 
See the footnote~15 in App.~A. 
Therefore, we do not need $\eta\pri$.}

The values of the C-type invariants are 
constant on each $G$-orbit as the A-type invariants.  
But we do not know whether they can be constructed 
by the fundamental fields and 
whether they have a geometric meaning, 
such as the moduli space of global symmetry in the case of A-types. 
We do not investigate these aspects in this paper, 
and concentrate on the B-type invariants.

\section{K\"{a}hler potentials $G$-invariant 
up to a K\"{a}hler transformation}

In the last section 
we discussed the strictly $G$-invariant K\"{a}hler potentials. 
In this section we discuss K\"{a}hler potentials $G$-invariant 
up to a K\"{a}hler transformation.
BKMU showed that they can be written in 
B-type K\"{a}hler potentials~\cite{BKMU}, 
which are generalizations of Zumino's one~\cite{Zu}. 
It is known that if there is no center in $\hat H$, 
B-type K\"{a}hler potentials are strictly $G$-invariant 
and are not independent of 
A- or C-type invariants~\cite{BKMU}. 
In the $O(N)$ model, this is the case and 
B-type K\"{a}hler potentials are strictly $G$-invariant.
Nevertheless, in this section 
we show that if and only if the orbit is open, 
does the K\"{a}hler manifold, $\GC/\hat H$, have 
a compact K\"{a}hler submanifold, $\GC/\tilde H$, 
and B-type K\"{a}hler potentials on $\GC/\hat H$ reduce 
to those on $\GC/\tilde H$, 
which is $G$-invariant up to a K\"{a}hler transformation.

\subsection{Closed orbits}
In this subsection, 
we consider the closed orbit. 
By using $\eta$ projections in Eq.~(\ref{closed-eta}), 
B-type K\"{a}hler potentials can be constructed 
as~\cite{BKMU,IKK}
\beq
 K_{{\rm B}i} = \log {\rm det}_{\eta_i} \xi\dagg\xi \;, 
 \label{closed-Btype}
\eeq
where ${\rm det}_{\eta_i}$ denotes a determinant 
in the $\eta_i$ projected matrix.\footnote{
Since, under a $g \in G$ transformation, 
the coset representative, $\xi$, 
is transformed as $\xi \gto \xi\pri = g \xi {\hat h}^{-1}$, 
where $\hat h \in \hat H$, 
$K_{{\rm B}i}$ is transformed as 
$K_{{\rm B}i} \gto K_{{\rm B}i} + 
\log {\rm det}_{\eta_i} {\hat h}^{-1\dagger} + 
\log {\rm det}_{\eta_i} {\hat h}^{-1}$. 
Here, we have used the projection conditions (\ref{proj}). 
Last two terms are changed to space-time total derivatives 
by the superspace integral. 
This can be understood as a K\"{a}hler transformation.} 
Since there is no center in $\hat H$, 
they are strictly $G$-invariant, 
contrary to the pure realization cases~\cite{BKMU,IKK}. 
Actually, 
\beq
 V_1 = {\rm det}_{\eta_1} \xi\dagg\xi 
     = \vec{v}\dagg \xi\dagg \xi \vec{v} / |\vec{v}|^2
\eeq
is not independent of the A-type invariant obtained in the last section. 
Moreover, $K_{\rm B2}$ is also expected not 
to be independent of the C-type invariant. 
(But we do not calculate an explicit form.)
Although the B-types are strictly $G$-invariant and 
are not independent of the A- or C-type invariants, 
we investigate these in more detail.

First we consider the case at the symmetric point. 
Since all broken generators are mixed-type and 
there is no pure-type multiplet, 
one can find no compact submanifold, 
which, if any, would be parameterized by pure-type multiplets.
Moreover, all broken generators belong to 
one representation, ${\bf N-1}$ of $H=O(N-1)$. 
Therefore, it is impossible to find any submanifold 
where some of the broken generators are changed to unbroken generators.

Secondly, we consider the non-symmetric points. 
At the non-symmetric points, 
the supersymmetric vacuum alignment occurs, 
and there appear $N-2$ pure-type multiplets and 
one mixed-type multiplet.
The pure-type multiplets belong to 
${\bf N-2}$ representation of $H\pri=O(N-2)$; 
on the other hand, 
the mixed-type multiplet is a singlet. 
Therefore, one may consider that there is a compact submanifold, 
but we show that this is not true. 

If there would be a compact submanifold, 
only one possibility is that 
it were a manifold parameterized by pure-type multiplets 
without one mixed-type multiplet. 
We, thus, change the mixed-type broken generator 
$Q \defeq X_{N-1}$ to an unbroken generator and 
define new unbroken and broken generators as 
\beq
  {\tilde {\cal H}} =\{ H_{\rm ss}, B_i, Q\},\;
  {\cal \GC}-{\tilde {\cal H}} =\{Z_i\} \;, 
\eeq
where $H_{\rm ss}$ is a Lie algebra of $H\pri =O(N-2)$ 
and $Z_i$ and $B_i$ are given in Eqs.~(\ref{ge.mixZ}) and (\ref{ge.mix}). 
Since $Q$ commutes with $H_{\rm ss}$, 
it is a center in $\{H_{\rm ss}, Q\}$. 
But, commutation relations of $Q$ with $Z_i$ and $B_i$ are 
\beq
 [Q,Z_i]=-i B_i,\;[Q,B_i] = i Z_i .\label{com-closed}
\eeq
Note that $Z_i$ and $B_i$ do not carry definite charges of $Q$. 
Moreover, ${\tilde {\cal H}}$ can not constitute a closed algebra. 
Thus, the submanifold $\GC/\tilde H$ 
can not be considered to be a coset space; 
but, since we can define it by using broken generators $Z_i$, 
we continue the argument.
The relation of coset representative $\xi$ of $\GC/\hat H\pri$ and 
the corresponding quantity (but not coset representative) 
$\zeta$ of $\GC/\tilde H$ is 
\beq
 \xi 
 &=& \exp (i \tilde\ph \cdot Z + i\Phi Q) \; \in \GC/\hat H\pri \non    
 &=& \exp (i \ph \cdot Z) \exp(i\Phi Q) \tilde h^{-1} \non
 &=& \zeta \exp(i\Phi Q) \tilde h^{-1}
\eeq
where $\ph^i$ in $\zeta \defeq \exp (i \ph \cdot Z)$ 
parameterize $\GC/\tilde H$. 
Here, $\tilde h \in \tilde H$ is needed, 
since the commutation relations of $Z_i$ and $Q$ include 
unbroken generators $B_i \in \tilde {\cal H}$ as Eq.~(\ref{com-closed}). 
If projections $\eta_i\pri = g_0 \eta_i {g_0}^{-1}$ on $\GC/\hat H\pri$ 
would satisfy $\eta\pri \tilde H \eta\pri =\tilde H \eta\pri$, 
they could be considered as also being projections on $\GC/\tilde H$, 
and the K\"{a}hler potentials (\ref{closed-Btype}) could reduce to 
K\"{a}hler potentials on $\GC/\tilde H$. 
But, unfortunately, they do not satisfy 
the projection conditions on $\GC/\tilde H$, Eq.~(\ref{proj.pure}).  
We, thus, conclude that 
K\"{a}hler potentials (\ref{closed-Btype}) on $\GC/\hat H$ 
do not reduce to those on $\GC/\tilde H$. 
In fact, the above argument at the non-symmetric points 
could be concluded by the arguments at the symmetric point: 
Even if the B-types reduced to those on any compact submanifold 
at the non-symmetric point, 
they could not connect to the symmetric point smoothly 
by any $\GC$-transformation. 
 
In the case of the closed orbit, 
we do not need B-type invariants, 
since they are not independent of A- or C-type invariants. 
In the next section we show that 
this feature is quite different in the open orbit.

\subsection{Open orbit}
In the open orbit, there is no vacuum alignment. 
There are $N-1$ pure-type generators and 
one mixed-type generator.
To consider a compact submanifold, 
as is done in the closed orbit, 
we change the mixed-type broken generator, 
$Q = X_{N-1}$, to an unbroken generator. 
New unbroken and broken generators are  
\beq
  {\tilde {\cal H}} =\{ H_{\rm ss}, B_i, Q\},\;
  {\cal \GC}-{\tilde {\cal H}} =\{Z_i\} \;,
\eeq
where $Z_i$ and $B_i$ are given in Eq.~(\ref{non-Her.open}).
Although $H_{\rm ss}$ and $Q$ are the same as in the closed-orbit case 
and $Q$ is a center in $\{H_{\rm ss}, Q\}$, 
non-Hermitian broken and unbroken generators, 
$Z_i$ and $B_i$, are different from the closed orbit. 
Namely, commutation relations between, $Z_i$ and $B_i$, and $Q_i$ are 
\beq
 [Q,Z_i]=Z_i,\;[Q,B_i] = - B_i \;. \label{com-open}
\eeq
Compare these equations with Eq.~(\ref{com-closed}). 
Different from the closed-orbit case, 
$Z_i$ and $B_i$ carry definite charges 
this time. From the second equation, 
$\tilde{\cal H}$ becomes an algebra and 
$\GC/\tilde H =O(N)^{\bf C}/O(N-2)^{\bf C}\times U(1)^{\bf C}\wedge B 
\simeq O(N)/O(N-2)\times U(1)$, 
where $B$ denotes the Borel group, is a coset space.
Since we obtain $[\tilde {\cal H},B_i]\sim B_j$, 
$B_i$ constitute a Borel subalgebra not only in $\hat{\cal H}$,
but also in $\tilde{\cal H}$.  
By using the first equation of (\ref{com-open}) and 
$[Z_i,Z_j]\sim Z_k$, 
the relation between the coset representative $\xi$ of $\GC/\hat H$ 
and the one $\zeta$ of compact homogeneous submanifold $\GC/\tilde H$ 
can be obtained as 
\beq
 \xi &=& \exp (i \tilde\ph \cdot Z + i\Phi Q) \; \in \GC/\hat H \non
     &=& \exp (i \ph \cdot Z) \exp(i\Phi Q) \non
     &=& \zeta \exp(i\Phi Q) \;. \label{rel.to-zeta}
\eeq
The coordinate transformation 
$\tilde\ph = \tilde\ph (\Phi,\ph)$ can be calculated
by using the Baker-Campbell-Hausdorff formula, 
but we do not need an explicit representation.
If we define 
\beq
 V = {\rm det}_{\eta} \xi\dagg\xi,\; 
\eeq
it is strictly $G$-invariant, 
since it is not independent of A-type invariant, $K_{\rm A}$, 
constructed in the last section: 
$V = \vec{v}\dagg \xi\dagg \xi \vec{v}/|\vec{v}|^2$. 
Therefore a B-type K\"{a}hler potential, 
\beq
 K_{\rm B} = c\log V = c\log {\rm det}_{\eta}\xi\dagg \xi,
\eeq
where $c$ is a real constant, is also strictly $G$-invariant.
This situation is the same as closed orbits. 
However, in this case, the $\eta$ projection on $\GC/\hat H$ also satisfies 
the projection conditions on the submanifold 
$\GC/\tilde H$, Eq.~(\ref{proj.pure}). 
Hence, by using Eq.~(\ref{rel.to-zeta}), $V$ can be calculated as
\beq
 V &=& {\rm det}_{\eta} \xi\dagg\xi 
    = {\rm det}\,(\eta \xi\dagg\xi\eta)\non
   &=& {\rm det}\,
        [\eta \exp(-i\Phi\dagg Q)\zeta\dagg\zeta \exp(i\Phi Q)\eta] \non
   &=& {\rm det}\,
    [\eta \exp(-i\Phi\dagg Q) \eta\zeta\dagg\zeta\eta \exp(i\Phi Q)\eta] \non
   &=& {\rm det}\,
    [(\eta \exp(-i\Phi\dagg Q) \eta)(\eta\zeta\dagg\zeta\eta)
     (\eta\exp(i\Phi Q)\eta)] \non
   &=& U \exp [i(\Phi\dagg - \Phi)]
\eeq
where $U \defeq {\rm det}_{\eta} \zeta\dagg\zeta$
is a corresponding quantity in $\GC/\tilde H$. 
Here, we have used the projection condition (\ref{proj.pure}) 
in the third line 
and a formula, $\log {\rm det}_{\eta} A = \tr (\eta \log A)$, 
and $\tr (\eta Q)=-1$ in the fourth line. From this equation, 
the K\"{a}hler potential can be calculated as 
\beq
 K_{\rm B} = c\log U + ic(\Phi\dagg - \Phi) \;.
\eeq 
The last term changes to the space-time total derivative 
by the superspace integral and 
$K_{\rm B}$ reduces to a K\"{a}hler potential $K_{\rm B0}$ 
of the compact submanifold $\GC/\tilde H$, 
\beq
 K_{\rm B0} 
     = c \log U 
     = c \log {\rm det}_{\eta} \zeta\dagg\zeta 
     = c \log \left(1 + |\ph|^2 + \1{4}\ph^2 \ph^{\dagger2}\right) .
  \label{B-com.}
\eeq
The last explicit form is calculated in App.~A.
Note that, although $K_{\rm B}$ is strictly $G$-invariant, 
$K_{\rm B0}$ is $G$-invariant up to a K\"{a}hler transformation.
$\GC/\tilde H$ is parameterized by scalar parts of pure-type multiplets 
without any mixed-type multiplet. 
This realization is called a pure realization. 
Hence, we conclude that the pure realization 
is embedded in the open orbit, but not in the closed orbit. 

In the open orbit, the most general effective 
K\"{a}hler potential can be written as a sum of 
the B-type K\"{a}hler potential, $K_{\rm B0}$ in Eq.~(\ref{B-com.}), 
on a compact submanifold $\GC/\tilde H \simeq O(N)/O(N-2)\times U(1)$, 
and an arbitrary function of strictly $G$-invariant, 
$K_{\rm A}$ in Eq.~(\ref{A-type}). 
Of course, if we choose the arbitrary function 
as $f(x)=c \log x$ in the latter, 
it becomes a sum of the former and the space-time total derivative. 
In this sense, the former can be included in the latter.  

\medskip
Before closing this section, 
we discuss a relation with the broken center models 
considered by Buchm\"{u}ller and Ellwanger~\cite{BE,El1,El2}. 
The pure realizations have the compact homogeneous 
target manifold $\GC/\tilde H$. 
In compact homogeneous K\"{a}hler manifolds, 
there exist a homeomorphism 
$ \GC/\tilde H \simeq G/H =G/H_{\rm ss} \times U(1)^n$. 
Here, $H_{\rm ss}$ is a semisimple subgroup of $\tilde H$ or $H$, 
and there are $n$ centers in $\tilde H$ or $H$. 
The K\"{a}hler potential is written as 
a sum of $n$ B-type K\"{a}hler potentials~\cite{IKK}. 
Buchm\"{u}ller and Ellwanger~\cite{BE} have considered 
models where $m (\leq n)$ centers in $\tilde {\cal H}$ are broken by hand 
from pure realizations. 
It was shown that $m$ linear combinations of 
B-type K\"{a}hler potentials are strictly $G$-invariant 
and that the K\"{a}hler potential can be written as a sum of 
$n$ B-type K\"{a}hler potentials and an arbitrary function of 
$m$ strictly $G$-invariants. 
Since these models are not pure realizations, 
they were expected to have linear model origins.  
However, it was not known whether they have linear model origins.

In the open orbit, we have shown that 
if one center in $\hat H$ is unbroken by hand, 
it reduces to the compact homogeneous manifold $O(N)/O(N-2)\times U(1)$.  
This can be considered as being an inverting procedure 
of Buchm\"{u}ller and Ellwanger. 
We, thus, have been able to find that 
special case (the case when one center is broken: $n=m=1$) of 
the broken-center models has an open orbit as a linear-model origin. 
A question as to whether the general cases of broken center models have 
linear model origins is discussed in the next section.

\section{Conclusions and Discussions}

The target spaces of the nonlinear sigma models, 
which have linear model origins, 
are obtained as $\GC$-orbits of the vacuum. 
In the $O(N)$ model, 
there are closed orbits and an open orbit, 
depending on the value of the $\GC$-invariant. 
(They are the orbit I and III in Fig.~1.) 
Both kinds of orbits can be written as complex coset spaces: 
$\GC / \hat H = O(N)^{\bf C}/O(N-1)^{\bf C}$ and 
$\GC / \hat H = O(N)^{\bf C}/O(N-2)^{\bf C} \wedge B$ respectively.
On the closed orbits, the vacuum alignment occurs (as Table~1), 
and the numbers of NG and QNG bosons change, 
with the total number remain unchanged, at various vacua. 
These two orbits are similar, except near to the origin, as in Fig.~1. 
Actually, the numbers of the pure-type multiplets and 
the mixed-type multiplets are the same 
at the non-symmetric points of the closed orbit and 
the generic points on the open orbit. 
(See second line of Table~1 and Table~2.) 
However, we have shown that, in the open orbit, 
the non-Hermitian unbroken generators, 
$B_i$ in Eqs.~(\ref{non-Her.open}) or (\ref{tr_basis}), 
constitute a Borel subalgebra in the complex isotropy $\hat {\cal H}$, 
but in the closed orbit, $B_i$ in Eq.~(\ref{ge.mix}) do not constitute it.
This difference plays a crucial role on the both orbits.

In the nonlinear realization with a linear model origin, 
the B-type K\"{a}hler potentials are strictly $G$-invariant and 
are not independent of A- or C-type invariants. 
To find a compact K\"{a}hler manifold as a submanifold, 
we have changed the mixed-type generator, $Q=X_{N-1}$, 
to an unbroken generator by hand in both orbits. 
In the closed orbit, 
$Z_i$ and $B_i$ does not carry definite charges of $Q$ 
as Eq.~(\ref{com-closed}); 
on the other hand, in the open orbit, 
$Z_i$ and $B_i$ carry definite $opposite$ 
charges of $Q$ as Eq.~(\ref{com-open}).
Moreover, in the closed orbit, 
the $\eta$-projection on the full manifold does not satisfy 
the perojection conditions (\ref{proj.pure}) on the compact submanifold; 
on the other hand, in the open orbit, 
it does satisfy the perojection conditions on the compact submanifold. 
We, thus, have found 
that any compact manifold is not embedded in the closed orbit, 
but it is embedded in the open orbit. From these differences, 
we have showed that 
the B-type K\"{a}hler potentials of the closed orbit 
are still just strictly $G$-invariant, 
and, on the other hand, 
that the B-type K\"{a}hler potential on the open orbit 
reduces to one of the compact homogeneous K\"{a}hler submanifold, 
$\GC / \tilde H = O(N)^{\bf C}/O(N-2)^{\bf C} \times U(1)^{\bf C}\wedge B 
\simeq O(N)/O(N-2) \times U(1)$, 
whose K\"{a}hler potential is 
$G$-invariant up to a K\"{a}hler transformation. 

It was known that a K\"{a}hler potential of a compact K\"{a}hler manifold 
can be written as a sum of B-type K\"{a}hler potentials. 
We may strengthen this property: 
It seems that even when the target manifold is non-compact, 
the B-type K\"{a}hler potentials 
can essentially live on compact K\"{a}hler manifolds 
(embedded in the target space). 
We can also say that B-type K\"{a}hler potentials 
can automatically find 
a compact homogeneous K\"{a}hler manifold as a submanifold 
in the full target manifold.

\medskip
Here we have discussions. 
These results can be generalized to more general models. 
First, we discuss closed orbits. 
It is known that a closed orbit has a symmetric point, 
such that a equation $\hat H = \HC$ is established. 
The maximal realization occurs at the symmetric point, 
there is no pure-type multiplet and 
one can not find any compact submanifold. 
Moreover, some mixed-type multiplets belong to non-singlet of $H$; 
hence, a combination with our previous results~\cite{Ni} leads to 
a supersymmetric vacuum alignment. 
The pure-type generators which arise by vacuum alignment 
are all non-Borelian, 
since the $\GC$-transformation does not change 
the commutation relations.
We, thus, can conclude that, on the closed orbits, 
there is no K\"{a}hler potential 
$G$-invariant up to a K\"{a}hler transformation.

Secondly, we discuss the open orbits. 
We do not know whether the open orbits have a vacuum alignment. 
However it does not seem that open orbits have 
a vacuum alignment, based on some examples 
(the open orbit of the $O(N)$ model in this paper and 
$U(N)$ with ${\bf N}$ in the previous paper~\cite{Ni}). 
If this is true, 
all pure-type generators constitute a Borel subalgebra in $\hat {\cal H}$. 
Since there is no vacuum alignment, from our previous results~\cite{Ni}, 
it can be concluded that all mixed-type generators are $H$-singlets. 
(Let its number to be $n$.)  
Hence, if we change these mixed-type generators to be unbroken, 
non-Hermitian unbroken generators constitute a Borel subalgebra in 
a complex isotropy which includes 
a complexification of $H_{\rm ss} \times U(1)^n$. 
(From now on we write $H$ as $H_{\rm ss}$.) 
But, to find a compact homogeneous manifold, 
we must change some off-diagonal charged 
broken generators to be unbroken. 
(This will show that there is no linear-model origin 
of the broken-center models 
considered by Buchm\"{u}ller and Ellwanger~\cite{BE,El1,El2}, 
$except$ $for$ the case where $one$ center is broken, 
considered in this paper.) 
This brings an arbitrariness of a choice of $\tilde H$, 
with $H_{\rm ss} \times U(1)^n$ being fixed. 
This is just a choice of 
a $G$-invariant complex structure~\cite{IKK,BFR,BN}. 
Namely, any compact homogeneous K\"{a}hler submanifold, 
$\GC/\tilde H \simeq G/H_{\rm ss}\times U(1)^n\; (n\geq 2)$, 
have such the arbitrariness~\cite{IKK,BFR,BN}.
Hence, it seems that 
we must add B-type K\"{a}hler potentials of 
$all$ compact K\"{a}hler submanifolds 
(even same real manifolds, 
$\GC / \tilde H \simeq G/H_{\rm ss}\times U(1)^n$, 
with different invariant complex structures) 
to its K\"{a}hler potential.  
We can show that 
they correspond to the choices of $\eta$ projections on $\GC / \hat H$. 
These aspects will be reported in 
the near future~\cite{Ni2}.

\medskip
Finally, we give some comments concerning to applications. 
Since we show that 
a pure realization is embedded in the open orbit, 
we can apply this phenomenon to 
seeking linear-model origins of 
pure realizations~\cite{IKK}--\cite{compact} 
by introducing a proper gauging. 
We hope that such investigation reaches to 
linear origins of the coset unification models~\cite{CUM}.

Since open orbits have pure-type multiplets anywhere, 
namely there exist no point such that $\hat H = \HC$ can be 
established and that a maximal realization can occur. 
Hence, if we gauge $G$, 
there is no point such that vector multiplets can get mass 
supersymmetrically by the supersymmetric Higgs mechanism. 
This brings about a spontaneous breaking of 
supersymmetry~\cite{gauging,Ku,WB}. 
These phenomena may be applied to 
dynamical supersymmetry breaking~\cite{SGT,DSB}.

\section*{Acknowledgments}
We thank K.~Higashijima for 
useful discussions and the reading of the manuscript.

\begin{appendix}
\section{K\"{a}hler potential on $O(N)/O(N-2)\times U(1)$}
Itoh, Kugo and Kunitomo have given the method to 
construct a K\"{a}hler potential of 
an arbitrary compact homogeneous K\"{a}hler manifold, 
$\GC/\tilde H \simeq G/H$~\cite{IKK}.
In this appendix, we construct a K\"{a}hler potential of 
$\GC/\tilde H = O(N)/O(N-2)\times U(1)$ by using their methods. 
We work in the changed basis (\ref{tr_basis}). 
Broken and unbroken generators are 
\beq
{\cal \GC}-\tilde{\cal H}=
 \left(
   \begin{array}{c|c|c}   
               &\LARGE{P}&         \\ \hline  
               &\LARGE{0}&\LARGE{P}\\ \hline  
               &         & \\ 
   \end{array}
 \right), 
\tilde{\cal H}=
 \left(
   \begin{array}{c|c|c}   
             Q &                                 & \\ \hline  
 \LARGE{\cal B}&\LARGE{{\cal H}_{\rm SS}^{\bf C}}& \\ \hline  
               &\LARGE{\cal B}                   & Q\\ 
   \end{array}
 \right) \;.
\eeq
A distinct point from the open orbit is that 
$Q=X_{N-1}$ is an unbroken generator. 
Compare thse equations with Eq.~(\ref{open-gene.}). 
Let a representative of the coset space 
$O(N)/O(N-2)\times U(1)$ as $\zeta$. From the equation
\beq
i\ph \cdot Z =
 \left(
   \begin{array}{c|c|c}   
               & i\ph^T  &     \\ \hline  
               &\LARGE{0}&-i\ph \\ \hline  
               &         &    \\ 
   \end{array}
 \right) \;,
\eeq
the representative $\zeta$ can be calculated explicitly as 
\beq
 \zeta = \exp(i\ph \cdot Z) = 
 \left(
   \begin{array}{c|c|c}   
             1 & i\ph^T  & {1\over 2} \ph^2 \\ \hline  
               &\LARGE{1}&-i\ph \\ \hline  
               &         &1    \\ 
   \end{array}
 \right)  \;.
\eeq
There exists one center $Q$ in ${\cal H}$ and 
broken generators carry positive charge: $[Q,Z_i]=Z_i$.\footnote{
For a convention, 
sines of charges carried by broken and unbroken generators are 
opposite to those in Ref.~\cite{IKK}. 
Therefore the projections are also opposite.}
The fundamental representation ${\bf N}$ can be decomposed as
\beq
 {\bf N} = {\bf 1}^1 \oplus ({\bf N-2})^0 \oplus {\bf 1}^{-1},
\eeq
in the $H$ representation, 
where the subscripts represent the charges under $Q$. 
Therefore, there is only one independent projection,\footnote{
Although 
$\eta\pri = 
 \left(
   \begin{array}{c|c|c}   
              0&         &     \\ \hline  
               &\LARGE{1}&\\ \hline  
               &         &1    \\ 
   \end{array}
 \right)$ also satisfy Eq.~(\ref{proj.pure}), 
it is not independent of $\eta$, 
since there is only one center in $H$. 
In the case of the pure realization, 
there exist independent projections 
as many as centers in $H$~\cite{IKK}.}
\beq
\eta=
 \left(
   \begin{array}{c|c|c}   
              0&         &     \\ \hline  
               &\LARGE{0}&\\ \hline  
               &         &1    \\ 
   \end{array}
 \right) , \label{eta_pure}
\eeq
which satisfies
\beq
 \eta\dagg = \eta\,,\; \eta \tilde H \eta =\tilde H \eta \,,\; \eta^2 = \eta.
  \label{proj.pure}
\eeq 
Note that $\eta$ in Eq.~(\ref{eta_pure}) coincides with 
$\eta$ on the open orbit, Eq.~(\ref{open-eta})
The second condition is satisfied, 
since all generators in $\tilde{\cal H}$ carry negative or zero charge: 
$[Q,B_i]=-B_i,\,[Q,H_{\ss}]=0$, 
where $H_{\ss}\in {\cal H}_{\rm ss}$~\cite{IKK}.
By using this projection, 
the K\"{a}hler potential can be calculated as~\cite{BKMU,IKK} 
\beq
 K = \log {\rm det}_{\eta} \zeta\dagg\zeta 
   = \log \left(1 + |\ph|^2 + \1{4}\ph^2 \ph^{\dagger2}\right) \;.
\eeq
This explicit form can be found in Refs.~\cite{El1,El2}.

\end{appendix}


\end{document}